\documentclass[aps, prb, twocolumn, superscriptaddress, floatfix ]{revtex4-2}
\usepackage{amsmath, amssymb, amsfonts}
\usepackage{inputenc}
\usepackage{graphicx, dcolumn, bm, color, epstopdf, hhline, mwe, ulem, xcolor, sidecap, subfloat, hyperref, array, tabularx, multirow}

\hypersetup{colorlinks=true, citecolor=blue, urlcolor=blue, linkcolor=blue}

\begin{document}

%\wideabs{

\preprint{APS/123-QED}

\title{Spin dynamics and magnetic excitations of quasi-1D spin chain Ca$_3$ZnMnO$_6$ }% Force line breaks with \\

\author{Suheon Lee} %(primary address Korea, conducted experiments, data evaluation, wrote draft)
\email[]{suheonlee@ibs.re.kr}
\affiliation{
Center for Artificial Low Dimensional Electronic System, Institute for Basic Science, Pohang 37673, Republic of Korea}

\author{D. T. Adroja}
\email[]{devashibhai.adroja@stfc.ac.uk}
\affiliation{
ISIS Neutron and Muon Source, STFC Rutherford Appleton Laboratory, Harwell Campus, Didcot, Oxfordshire OX11 0QX, United Kingdom}
\affiliation{Highly Correlated Matter Research Group, Physics Department, University of Johannesburg, Auckland Park 2006, South Africa}

\author{Qing Zhang}
\affiliation{ School of Physics, Shandong University, Jinan 250100, China}
\author{Gheorghe Lucian  Pascut}
\email[]
{glucian.pascut@usm.ro}
\affiliation{MANSID Research Center, University Stefan Cel Mare, 720229 Suceava, Romania}
\author{Kristjan Haule}
\affiliation{Department of Physics, Rutgers University, Piscataway, New Jersey 08854, USA}
\author{A.D. Hillier}
\affiliation{
ISIS Neutron and Muon Source, STFC Rutherford Appleton Laboratory, Harwell Campus, Didcot, Oxfordshire OX11 0QX, United Kingdom}
\author{M. Telling}
\affiliation{
ISIS Neutron and Muon Source, STFC Rutherford Appleton Laboratory, Harwell Campus, Didcot, Oxfordshire OX11 0QX, United Kingdom}
\author{W. Kockelmann}
\affiliation{
ISIS Neutron and Muon Source, STFC Rutherford Appleton Laboratory, Harwell Campus, Didcot, Oxfordshire OX11 0QX, United Kingdom}

\author{Sang-Wook Cheong}
\affiliation{Department of Physics, Rutgers University, Piscataway, New Jersey 08854, USA}
\affiliation{Keck Center for Quantum Magnetism and Department of Physics and Astronomy,
Rutgers University, Piscataway, NJ, 08854, USA}
\author{K.-Y. Choi}
\affiliation{Department of Physics, Sungkyunkwan University, Suwon 16419, Republic of Korea}

\begin{abstract}
To reveal the structure-property relationship in quasi-one-dimensional (1D) spin-chain system Ca$_3$ZnMnO$_6$, we present comprehensive results, combining basic physical characterizations such as muon spin relaxation/rotation ($\mu$SR), neutron powder diffraction (NPD), inelastic neutron scattering (INS), and theoretical calculations. Ca$_3$ZnMnO$_6$ features a dominant intrachain coupling $J_1$ and two distinct interchain interactions $J_2$ and $J_3$, and it undergoes antiferromagnetic ordering below $T_{\mathrm{N}}=25$~K, as revealed by dc magnetic susceptibility and specific-heat measurements. Zero-field $\mu$SR shows persistent spin dynamics below $T_{\mathrm{N}}$, suggesting unconventional magnetic excitations in the ordered state. NPD results indicate a commensurate magnetic ground state with a propagation vector $\mathbf{k}=0$, where the Mn spins lie in the $ab$-plane. INS spectra display dispersive magnetic excitations extending up to about 5~meV, with an energy gap smaller than 0.5~meV. Notably, these spectra exhibit three-dimensional (3D) gapped features rather than the expected 1D behavior, yet spin-wave dispersion analysis confirms an underlying quasi-1D energy hierarchy. We discuss this apparent paradox of 3D-like magnetic excitations in a quasi-1D system in terms of the energy hierarchy modified by nonmagnetic-ion substitution and finite-temperature first-principles calculations. We also suggest that Ca$_3$ZnMnO$_6$ could be a potential candidate for an M-type altermagnet. 
\end{abstract}

\maketitle

\section{\label{sec1}Introduction}

During recent decades, much effort has been devoted to understanding one-dimensional (1D) spin systems. 1D spin chains host a rich variety of ground states with exotic phenomena, such as gapped Haldane excitations, spinon confinement, Bose-Einstein condensation, and spin-Peierls transitions~\cite{Haldane1980,Haldane1982,Haldane1985,Ruegg,Lake,Hase}. It is well known that the ideal 1D spin chain evades long-range magnetic order at finite temperatures due to enhanced quantum fluctuations arising from reduced dimensionality~\cite{Bethe}. However, perturbations, including weak interchain interactions and nonmagnetic impurities, can give rise to long-range ordering at low temperatures~\cite{Gomez-Santos,Uchiyama}. Particularly, in the presence of interchain couplings, spin dynamics are determined by energy hierarchy between intra- and interchain interactions. 

\begin{figure}[h]
	\centering
	\includegraphics[width=1\linewidth, trim={0cm 80 0 50},clip]{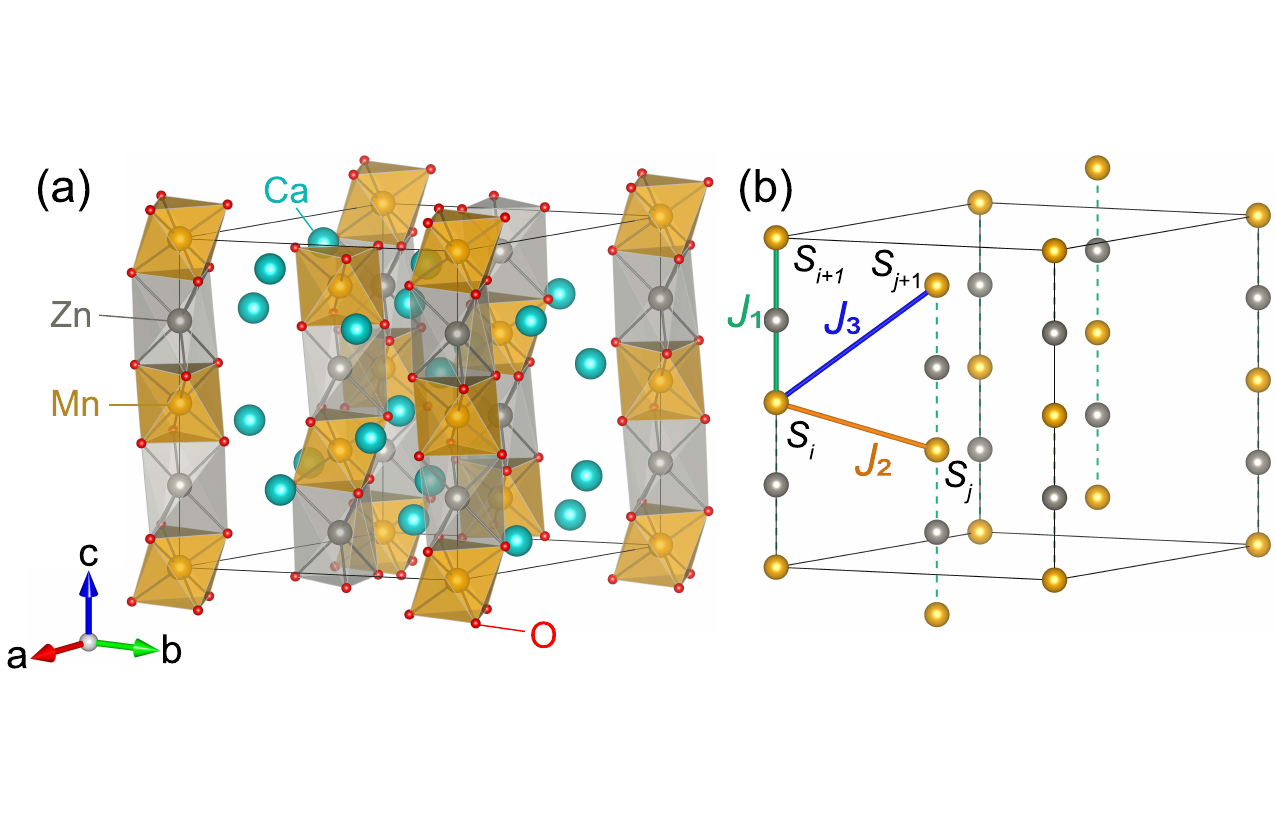}
	\caption{\textbf{Schematic structure of Ca$_3$ZnMnO$_6$.} (a) Crystal structure of Ca$_3$ZnMnO$_6$. The cyan, grey, yellow, and red spheres represent Ca, Zn, Mn, and O atoms, respectively. The Zn atoms are on the trigonal prismatic site, and the Mn atoms are on the octahedral site. (b) Spin topology of Ca$_3$ZnMnO$_6$. The Mn$^{4+}$ ions interact with each other through Mn-O-O-Mn exchange pathways along the \textit{c}-axis, $J_1$ (intra-chain Mn-Mn distance $d_\textrm{Mn-Mn}=5.32$~\AA). The Mn chains are coupled via two different interchain exchange interactions $J_2$ ($d_\textrm{Mn-Mn}=5.57$~\AA) and $J_3$ ($d_\textrm{Mn-Mn}=6.36$~\AA). The Ca, Zn, and O atoms are omitted for clarity.}
	\label{Figure1}
\end{figure}

Recently, a new family of the spin-chain systems A$_3$BB'O$_6$ (A = Ca, Sr; B, B' = transition metals) has been discovered. A$_3$BB$'$O$_6$ crystallizes in the K$_4$CdCl$_6$-type structure (space group $R\bar{3}c$), in which B and B$'$ ions constitute the ferromagnetic spin chains along the \textit{c}-axis as shown in Fig.~\ref{Figure1}(a) ~\cite{NGUYEN1995300}. The spin chains are antiferromagnetically coupled and form a distorted triangular network, giving rise to geometrical frustration. Furthermore, this system provides fertile ground to explore magnetic excitations in both half-integer and integer spin systems due to structural stability and the diverse combinations with different magnetic ($3d$, $4d$, and $5d$) and nonmagnetic ions. The combined effects of low dimensionality and geometrical frustration have attracted intensive attention from both theoretical and experimental investigations~\cite{Nguyen,Flahaut2003,Wu2005, Mohapatra2007, Agrestini2008, Sarkar2010, Yin2013, Jain2013, Lefrancois2014, Hillier2011, Ou2014, Toth2016, McClarty2020}.

Among the series compounds, Ca$_3$ZnMnO$_6$ shows slightly different exchange topology [Fig.~\ref{Figure1}(b)]~\cite{Kawasaki}. The substitution of nonmagnetic ion (Zn$^{2+}$) into B-sites introduces an antiferromagnetic (AFM) exchange interaction $J_1$ between Mn$^{4+}$ ($S=3/2$) in the spin chain along the \textit{c}-axis. The Mn spin chains interact with each other via two different interchain couplings $J_2$ and $J_3$. The magnetic susceptibility exhibits an AFM ordering at $T_\textrm{N} = 25$~K with a divergence between zero-field-cooled and field-cooled data~\cite{Kawasaki,Ruan}. The Curie-Weiss fits provide the Curie-Weiss temperature of $\Theta_\textrm{CW} = -47$~K. High-field magnetization data manifest no magnetization plateau up to 30~T, suggesting small anisotropy as theoretically predicted for the $S = 3/2$ spin chain system~\cite{Ruan,Kitazawa}. Noticeably, the high-frequency electron spin resonance (ESR) experiments reveal the presence of the zero-field spin gap $\Delta = 166$~GHz (0.69~meV), originating from an easy-plane anisotropy and exchange interaction rather than Ising anisotropy or Dzyaloshinsky-Moriya interaction~~\cite{Ruan}. Taken together, the small easy-plane anisotropy of Ca$_3$ZnMnO$_6$ plays a significant role in the antiferromagnetically ordered state.

The previous theoretical study using density functional theory (DFT) demonstrates that not only intrachain but also interchain interactions are antiferromagnetic in Ca$_3$ZnMnO$_6$ ~\cite{Chakraborty}. The AFM intrachain coupling $J_1$ opposes the typical ferromagnetic exchange interaction observed along the chain direction in A$_3$BB'O$_6$. The AFM exchange coupling constants are calculated to be $J_1 = 2.46$~meV, $J_2 = 1.76$~meV, and $J_3 = 0.34$~meV~\cite{Chakraborty}. The comparable values of $J_1$ and $J_2$ suggest the possibility that Ca$_3$ZnMnO$_6$ host three-dimensional (3D) magnetism rather than 1D magnetism. Furthermore, the DFT calculations, including the spin-orbit coupling, deduce a tiny value of the magnetocrystalline anisotropy about 0.01~meV. However, the spin dynamics and inherent magnetic excitations in Ca$_3$ZnMnO$_6$ remain elusive.

In this paper, we present basic characterizations, $\mu$SR, NPD, and INS results of the quasi-1D spin chain system Ca$_3$ZnMnO$_6$. Contrary to the expected exchange topology of a quasi-1D spin chain, we observe the characteristics of 3D magnetism, such as the absence of high-$T$ broad maximum in the specific heat data and the dispersive magnetic excitations, rather than those for 1D spin systems. Nevertheless, our linear spin-wave calculations suggest that the energy hierarchy of exchange interactions corresponds to the quasi-1D spin chain. Such disagreement in dimensionality can be understood in terms of the enhanced interchain coupling constant by nonmagnetic substitution and unconventional superexchanges with
orbital selectivity found from first-principles calculations.

\section{\label{sec2}Experimental Details}
%Sample synthesis
Polycrystalline samples of Ca$_3$ZnMnO$_6$ were synthesized using a solid-state reaction method as described in Ref.~\cite{Kawasaki}. Stoichiometric amounts of high purity ($> 99.9$~\%) SrCO$_3$, ZnO and MnO$_2$ powders were carefully mixed and ground. The mixtures were calcined at  900~$^\circ$C for 24~h and subsequently sintered at 1200~$^\circ$C for $24-36$~h with several intermediate steps of grinding. The quality of the specimen was confirmed by powder X-ray diffraction (XRD) experiments at room temperature. dc magnetic susceptibility was measured using a commercial SQUID magnetometer (MPMS, Quantum Design) in the temperature range of $T=2-300$~K. Specific heat measurements were carried out using a physical property measurement system (PPMS, Quantum Design) with the conventional relaxation method in the applied fields of 0 and 7~T.

Zero-field (ZF) muon spin relaxation ($\mu$SR) experiments were conducted at the MuSR spectrometer at the ISIS Neutron and Muon Source (Didcot, UK, RB1010488). Polycrystalline Ca$_3$ZnMnO$_6$ were finely ground and then packaged using an alumina foil. ZF-$\mu$SR experiments were performed using an Oxford $^4$He cryostat in the temperature range of $T=1.4-170$~K. All the $\mu$SR spectra were fitted using the Mantid software~\cite{Arnold}. 

Neutron powder diffraction (NPD) experiments (RB920526) were carried out using the GEM diffractometer at the ISIS Neutron and Muon Source (Didcot, UK). The powder sample was filled into a thin-walled, cylindrical  vanadium container of 8mm diameter. Data were collected between 5 K and 30 K using a He-flow cryostat.  The obtained diffraction patterns of Ca$_3$ZnMnO$_6$ were analyzed by Rietveld refinement using the FullProf software package~\cite{Carvajal}. In the antiferromagnetically ordered state, we analyzed the magnetic symmetry using the BasIreps software embedded in the FullProf for determining the Mn sublattice that occupies the Wyckoff position 6\textit{b}. 

Inelastic neutron scattering (INS) measurements were performed on the MARI (RB1320246 and RB1419655
%\textcolor{red}{RB1320246 and RB1419655}
) and IRIS (RB920482
%\textcolor{red}{RB920482}
) spectrometers at the ISIS Neutron and Muon Source (Didcot, UK). For the direct geometry MARI spectrometer, the finely ground powder ($\sim5$~g) was packed with an alumina foil contained in an alumina can and then inserted into a closed-cycle cryostat. The energy of the incident neutrons ($E_\textrm{i} = 12.8$~meV) was selected using the Gd Fermi chopper rotating at 100~Hz, resulting in a $\sim 0.5$~meV full width at half maximum (FWHM) energy resolution of the elastic scattering. Further, to elucidate low-lying magnetic excitations, INS experiments were performed at the high-resolution time-of-flight IRIS spectrometer. The indirect geometry IRIS spectrometer was operated with PG002 analyzers, providing the analyzing energy of $E_\textrm{f} = 1.84$~meV and a FWHM energy resolution of 17.5~$\mu$eV at the elastic position. The sample measured on IRIS was the identical powder sample used for the MARI experiment. The observed magnetic excitations were calculated using SpinW software package based on linear spin wave theory~\cite{Toth}. 

\section{\label{sec3}Results and Discussion}

\subsection{Magnetic susceptibility and specific heat}

\begin{figure}[h]
	\centering
	\includegraphics[width=1\linewidth, trim={0cm 350 0 370},clip]{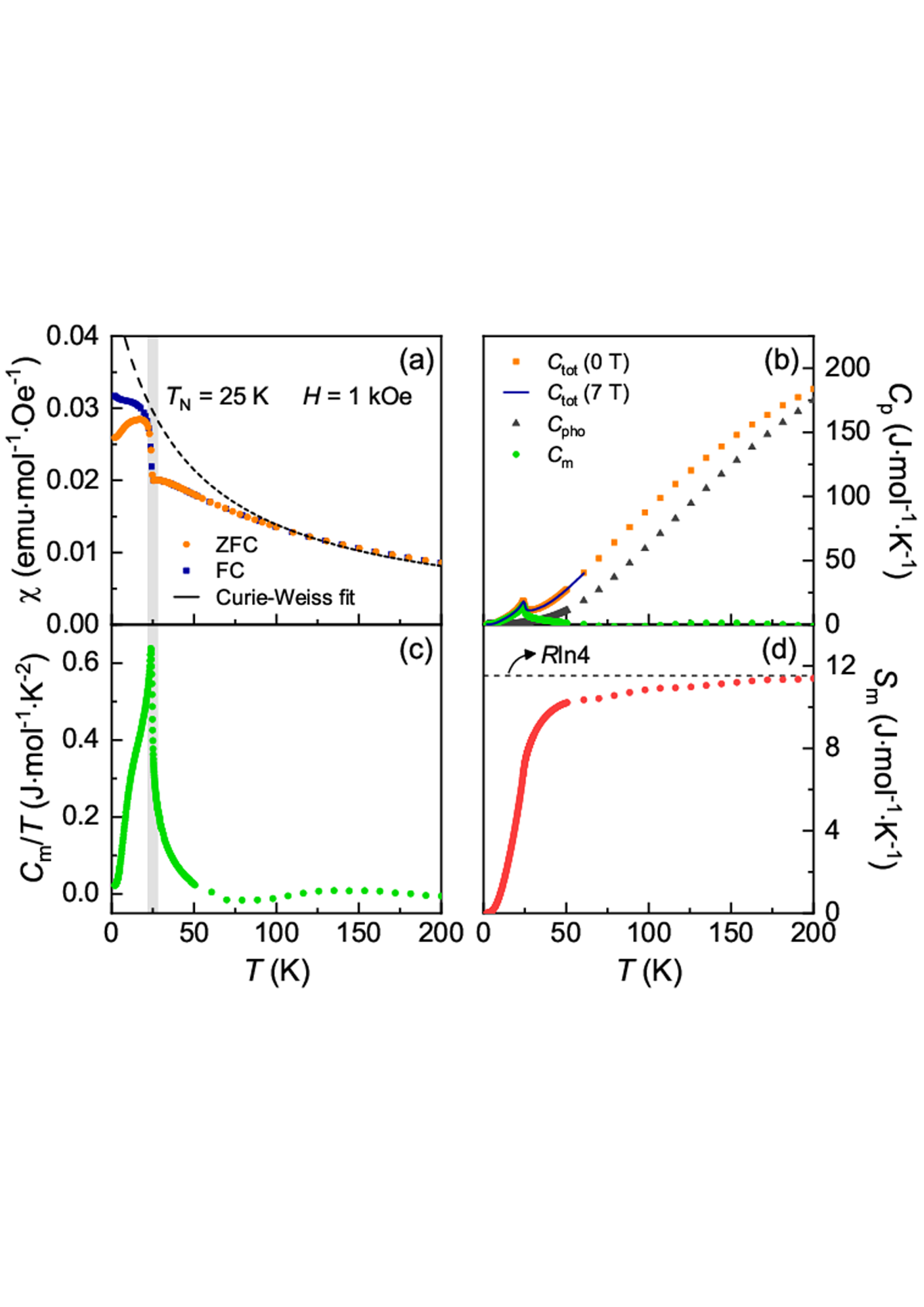}
	\caption{\textbf{Magnetic and thermodynamic characterization results of Ca$_3$ZnMnO$_6$.} (a) Temperature dependence of the dc magnetic susceptibility measured in $H=1$~kOe. The shaded region indicates the antiferromagnetic transition at $T_\textrm{N}=25$~K. The dashed curve denotes a Curie-Weiss fit to the high-temperature magnetic susceptibility data. (b) Specific heat as a function of temperature in two representative magnetic fields (0 and 7~T) and its decomposition into phonon and magnetic contributions. (c) Temperature dependence of the magnetic specific heat divided by temperature $C_\textrm{m}/T$ in zero field. (d) The magnetic entropy of Ca$_3$ZnMnO$_6$ as a function of temperature in zero field. The dashed line indicates the theoretical value of the magnetic entropy $S_\textrm{m} = R\textrm{ln}4 = 11.52$~J$\cdot$mol$^{-1}\cdot$K$^{-1}$.}
	\label{Figure2}
\end{figure}

Figure~\ref{Figure2}(a) displays the temperature dependence of the dc magnetic susceptibility measured in $H = 1$~kOe. With decreasing temperature, $\chi(T)$ exhibits a broad maximum at $T = 30$~K, followed by an antiferromagnetic transition at $T_\textrm{N} = 25$~K. The Curie-Weiss fits to the data above 150~K provide the Curie constant $C = 1.96(4)$~emu$\cdot$K$\cdot$mol$^{-1}\cdot$Oe$^{-1}$ and the Weiss temperature $\Theta_\textrm{CW} = -41(5)$~K. The effective magnetic moment was evaluated to be $\mu_\textrm{eff} = 3.96(5)~\mu_\textrm{B}$, comparable to the theoretical value for the Mn$^{4+}$ moments (3$d^3$, $S=3/2$; 3.87~$\mu_\textrm{eff}$). The obtained parameters are in agreement with the previously reported values~\cite{Kawasaki, Ruan}. The deviation of $\chi(T)$ from the Curie-Weiss law below 100~K is attributed to the development of magnetic correlations for $T>|\Theta_\textrm{CW}|$, which is characteristic of frustrated spin systems. 

The temperature dependence of the specific heat is depicted in Fig.~\ref{Figure2}(b). As the temperature is lowered, the specific heat gradually decreases and shows a $\lambda$-like peak at $T_\textrm{N}=24$~K, consistent with the AFM ordering observed in $\chi(T)$. To single out the magnetic contribution ($C_\textrm{m}$) to the total specific heat ($C_\textrm{tot}$), we estimate the lattice contribution ($C_\textrm{pho}$) using the sum of a Debye model and an Einstein term,

\begin{equation}
\begin{aligned}
C_\textrm{pho}(T) & = 9 C_\textrm{D} Nk_\textrm{B}\Big(\frac{T}{\theta_\textrm{D}}\Big)^3\int^{\theta_\textrm{D}/T}_{0}\frac{x^4e^x}{(e^x-1)^2}dx  \\ 
	& + 3Nk_\textrm{B}C_\textrm{E} \Big(\frac{\theta_\textrm{E}}{T}\Big)^2\frac{e^{(\theta_\textrm{E}/T)}}{(e^{(\theta_\textrm{E}/T)}-1)^2}.
\end{aligned}
\end{equation}
Here, $N$ is the Avogadro number, $k_\textrm{B}$ is the Boltzmann constant, $C_\textrm{D}$ and $C_\textrm{E}$ are the coefficients for distinct atoms in the crystal, and $\theta_\textrm{D}$ and $\theta_\textrm{E}$ are the Debye and Einstein temperatures. We obtain the best fits for $T>50$~K with the parameters of $C_\textrm{D}=6.1(9)$, $\theta_\textrm{D}=379(11)$~K, $C_\textrm{E}=5.5(7)$, and $\theta_\textrm{E}=790(19)$~K. The sum of the coefficients ($C_\textrm{D} + C_\textrm{E} = 11.6$) is comparable with the total number of atoms per the formula unit divided by the number of Mn ions. After subtracting the lattice contribution ($C_\textrm{pho}$) from the total specific heat ($C_\textrm{tot}$), we obtain the temperature dependence of the magnetic specific heat ($C_\textrm{m}$), displaying a sharp peak at $T_\textrm{N}$ = 25~K [green circles in Fig.~\ref{Figure2}(b)]. In addition, we observe a shoulder feature extending to 60~K, which likewise appears as a broad hump in $\chi(T)$ within the same $T$ range. These findings suggest a weak feature pertinent to short-range correlations above $T_\textrm{N}$. We note that the total specific heat shows no magnetic field dependence up to 7~T, ruling out the possibility of spin freezing in the magnetically ordered state. 

By integrating $C_\textrm{m}(T)/T$, we evaluate the magnetic entropy as a function of temperature [Fig.~\ref{Figure2}(d)]. With increasing temperature, $S_\textrm{m}$ shows a steep increment, and the slope becomes somewhat moderate through 45~K. This temperature is similar to the Curie-Weiss temperature ($|\Theta_\textrm{CW}|=41$~K) extracted from the dc magnetic susceptibility. At higher temperatures, $S_\textrm{m}$ is fully recovered to the theoretically expected value for $S = 3/2$, $S_\textrm{m}^\textrm{theory} = R\textrm{ln}4 = 11.52$~J$\cdot$mol$^{-1}\cdot$K$^{-1}$ at 200~K.

\subsection{Muon spin relaxation}

\begin{figure}[h]
	\centering
	\includegraphics[width=1\linewidth, trim={0cm 50 0 0},clip]{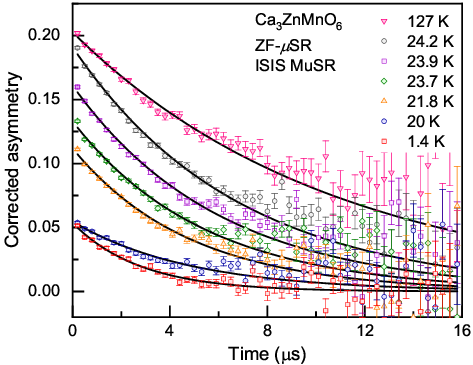}
	\caption{\textbf{ZF-$\mu$SR spectra of Ca$_3$ZnMnO$_6$.} Representative ZF-$\mu$SR spectra at various temperatures. The time-independent background is subtracted from the total asymmetry. The solid lines denote the fits to the data.}
	\label{Figure3}
\end{figure}

We performed ZF-$\mu$SR experiments of Ca$_3$ZnMnO$_6$ to investigate the evolution of spin correlations. The temperature dependence of ZF-$\mu$SR spectra is depicted in Fig.~\ref{Figure3}(a). In the paramagnetic state, the asymmetry exponentially decays in time, originating from rapidly fluctuating magnetic moments. As the temperature is lowered, the muon spin depolarization becomes faster and shows no coherent muon spin oscillation signal at low temperatures. Rather, we find the loss of initial asymmetry below $T_\textrm{N}=24$~K, indicating the occurrence of long-range magnetic order. The initial asymmetry loss is attributed to strong damping in the antiferromagntically ordered state at an early time, which is beyond the time resolution of the pulsed muon source. Below $T_\textrm{N}$, the residual asymmetry is fully depolarized at a longer time, suggesting the presence of dynamically fluctuating moments even in the antiferromagnetically ordered state. 

\begin{figure}[h]
	\centering
	\includegraphics[width=1\linewidth, trim={0cm 50 80 50},clip]{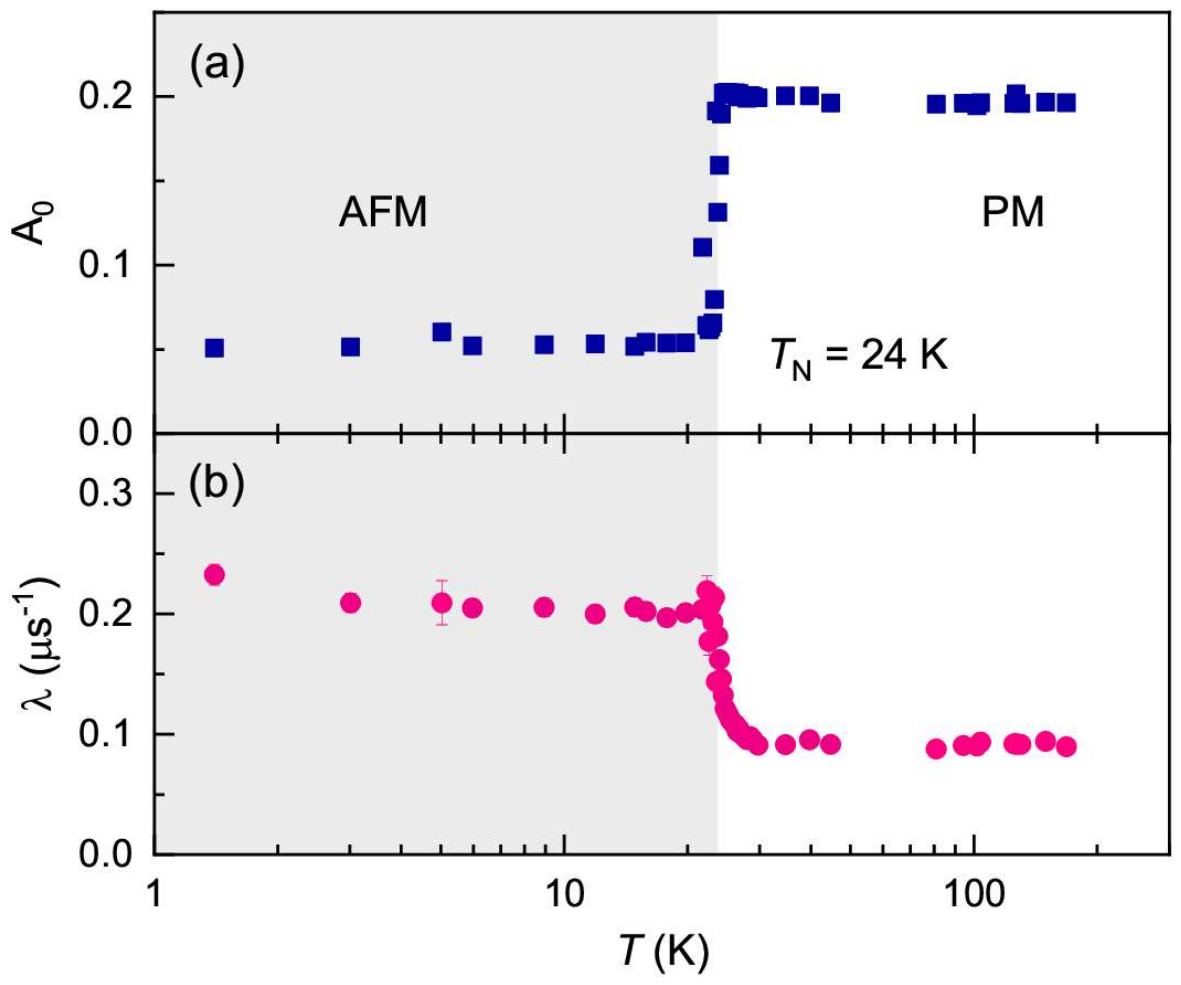}
	\caption{\textbf{ZF-$\mu$SR parameters of Ca$_3$ZnMnO$_6$.} (a) Temperature dependence of the initial asymmetry. (b) Muon spin relaxation rate as a function of temperature. The shaded region represents the antiferromagnetically ordered state below $T_\textrm{N} = 24$~K.}
	\label{Figure4}
\end{figure}

For a quantitative analysis, the ZF-$\mu$SR spectra were fitted with the sum of a simple exponential function and a temperature-independent constant background $A_z(t) = A_0\textrm{exp}(-\lambda t) + A_\textrm{bkg}$. Here, $\lambda$ is the muon spin relaxation rate, $A_0$ is the initial asymmetry, and $A_\textrm{bkg}$ is the temperature-independent background. In Fig.~\ref{Figure4}, we plotted the temperature dependence of the obtained parameters in a semilog scale. As displayed in Fig.~\ref{Figure4}(a), $A_0$ drops rapidly through $T_\textrm{N}$ to approximately 1/3 of its high-$T$ value, suggesting the entrance into the antiferromagnetically ordered state of full sample volume. 

On the other hand, the muon spin relaxation rate shows a peculiar thermal evolution [Fig.~\ref{Figure4}(b)]. In the high-$T$ paramagnetic phase, $\lambda$ exhibits a temperature-independent behavior, supporting that the exchange fluctuations of the Mn spins predominantly govern the muon spin depolarization. At high $T$, we estimate the exchange coupling constant $J=1.03$~meV from $k_\textrm{B}\Theta_\textrm{CW} = 2zS(S+1)J/3$ with the coordination number of the nearest neighbor $z=2$, providing the exchange fluctuation rate $\nu = \sqrt{z}JS/\hbar \sim 3.3 \times 10^{12}$~s$^{-1}$. By using the relation $\lambda = 2\Delta^2/\nu$ in the exchange narrowing limit~\cite{Yaouanc2011}, the internal field distribution width is deduced to be $\Delta/\gamma_\mu = 4.5$~kG with $\lambda(T =170$~K$) = 0.19$~$\mu$s$^{-1}$. With decreasing temperature, $\lambda$ steeply increases due to the development of magnetic correlations between the Mn spins and then flattens out below $T_\textrm{N}$. The leveling off of the muon spin relaxation at low-$T$ is so-called persistent spin dynamics, widely reported in a range of frustrated spin systems, such as quantum spin ice, quantum spin liquids, and quantum and weakly-ordered magnets~\cite{Mendels,Dunsiger,Yaouanc2015}. The observed persistent spin dynamics in Ca$_3$ZnMnO$_6$ suggests the presence of unconventional low-energy magnetic excitations in the ground state as in other weakly-ordered systems~\cite{Yaouanc2015,Xu,Reotier}.

\subsection{Neutron powder diffraction}

\begin{figure}[h]
	\centering
	\includegraphics[width=1\linewidth]{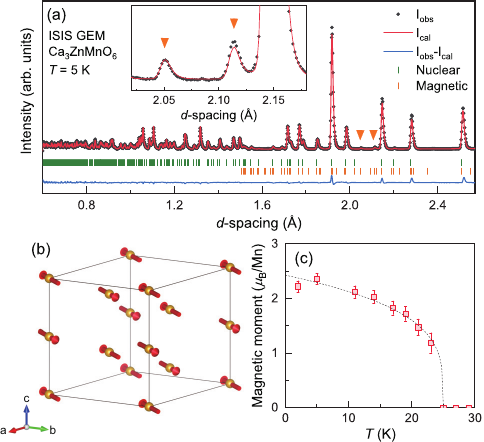}
	\caption{\textbf{Neutron powder diffraction results of Ca$_3$ZnMnO$_6$.} (a) Neutron powder diffraction pattern of Ca$_3$ZnMnO$_6$ at $T = 5$~K. The refinement results are superimposed on the observed data as a solid red curve. Vertical ticks indicate calculated nuclear and magnetic Bragg peak positions. The difference between the data and calculation is depicted below the Bragg peak positions. The orange downward triangles denote the magnetic Bragg peaks below $T_\textrm{N}$. The inset shows the enlarged view of the magnetic Bragg peaks. (b) Magnetic structure of Ca$_3$ZnMnO$_6$ determined by the Rietveld refinement at $T=5$~K. (c) Temperature dependence of the magnetic moment for Mn$^{4+}$ ($S = 3/2$) ions. The dashed curve denotes a power-law fit. }
	\label{Figure5}
\end{figure}
 
We further carried out NPD measurements of Ca$_3$ZnMnO$_6$ to determine the magnetic ordering pattern. The NPD results of Ca$_3$ZnMnO$_6$ are summarized in Fig.~\ref{Figure5}. For $T>T_\textrm{N}$, the observed NPD pattern is well described by the nuclear Bragg peaks with the reported space group ($R\bar{3}c$) and similar values of parameters (Table~\ref{Table1})~\cite{Kawasaki,Ruan}. As the temperature was lowered below $T_\textrm{N}$, the magnetic Bragg peaks began to appear, signifying long-range magnetic ordering. In particular, we observed two magnetic Bragg peaks at around 2~\AA, as highlighted by the inverse triangles in Fig.~\ref{Figure5}(a). All of the magnetic reflections below $T_\textrm{N}$ are in agreement with a propagation vector of $\bm{k}=0$ and magnetic symmetry ($R\bar{3}$). 

\begin{table}
\caption{Structural parameters of Ca$_3$ZnMnO$_6$ obtained from the neutron diffraction analysis at 25~K for space group $R\bar{3}c$ (No. 167). Lattice parameters are $a=9.133(1)$~\AA~and $c=10.623(2)$~\AA. Bragg R-factor and Rf-factor are 4.72 and 4.06, respectively.} 
\begin{ruledtabular}
\begin{tabular}{lllll}\label{Table1}
Atoms & Site & $x$ & $y$  & $z$\\
\hline
Ca & 18e & 0.363(2) & 0 & 0.250 \\
Zn & 6a & 0 & 0 & 0.250 \\
Mn & 6b & 0 & 0 & 0 \\
O & 36f & 0.179(1) & 0.025(1) & 0.107(1) \\
\end{tabular}
\end{ruledtabular}
\end{table}

From Rietveld refinements using the propagation vector \textit{\textbf{k}}, we reveal the temperature dependence of the magnetic moment for the Mn spins and the magnetic ground state [Figs.~\ref{Figure5}(b) and \ref{Figure5}(c)]. The refinement results show a good agreement with the experimental data with Bragg R-factor = 4.51, Rf-factor = 3.73, and magnetic R-factor = 27.9. The magnetic structure of Ca$_3$ZnMnO$_6$ exhibits an AFM arrrangement of the Mn spins with 2.36(9)~$\mu_\textrm{B}$ in the $ab$-plane, while the orientation of the Mn spins with regard to the $a$- and $b$-axes cannot be determined due to powder averaging. Our NPD experiments determined the magnetic structure below $T_\textrm{N}$, however like in the previous NPD study~\cite{Kawasaki}, the Mn spin direction in the \textit{ab}-plane remains elusive. 

The thermal evolution of the magnetic moments below $T_\textrm{N}$ is well described by the power-law $\mu_\textrm{Mn}(T) = \mu_\textrm{Mn}(0)(1-T/T_\textrm{N})^\beta$, yielding $\mu_\textrm{Mn}(0) = 2.43(6)$~$\mu_\textrm{B}$ and $\beta = 0.26(2)$. The extracted exponent $\beta$ is incompatible with the expected values based on the criticality theory [two-dimensional Ising ($\beta=0.125$), 3D Ising ($\beta=0.326$), 3D Heisenberg ($\beta=0.367$), and 3D XY ($\beta=0.35$) spins]. Such discrepancy indicates a deviation from the Ising nature in Ca$_3$ZnMnO$_6$, supported by the high-frequency ESR measurements~\cite{Ruan}. Despite the presence of three different spin chains in the unit cell, the magnetic order below $T_\textrm{N}$ and the absence of the Ising character imply the significant role of interchain interactions in the spin topology.

\subsection{Inelastic neutron scattering}

\begin{figure}[h]
	\centering
	\includegraphics[width=1\linewidth,trim={0cm 130 0 100},clip]{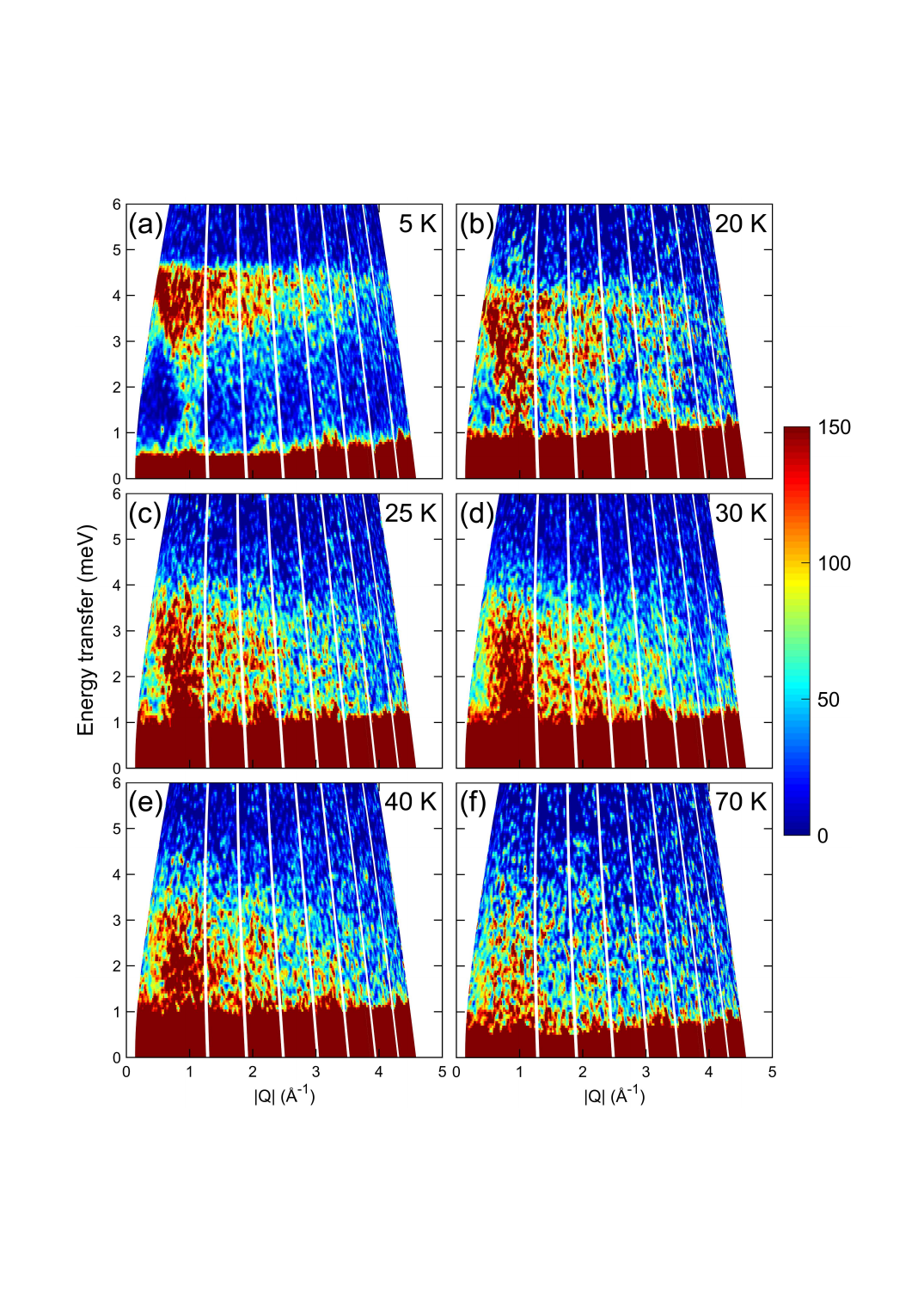}
	\caption{\textbf{Intensity of the spin excitations at various temperatures.} Color-coded inelastic neutron scattering intensity maps with $E_\textrm{i}=12.8$~meV at different temperatures (a) 5~K, (b) 20~K, (c) 25~K, (d) 30~K, (e) 40~K, and (f) 70~K.}
	\label{Figure6}
\end{figure}

To investigate the magnetic excitations of Ca$_3$ZnMnO$_6$, we conducted inelastic neutron scattering experiments as depicted in Fig.~\ref{Figure6}. As expected from the magnetic susceptibility and magnetic specific heat results, the inelastic neutron scattering spectra at $T=70$~K show diffusive magnetic excitations observed due to the development of short-range magnetic correlations. As the temperature is lowered, the magnetic excitations gradually develop at around $|Q| = 1$~\AA$^{-1}$~for $T>T_\textrm{N}$. Upon further cooling, we observe the spin-wave dispersion below $T_\textrm{N}$ featured by a broad flat band at around 4.2~meV. Taking into account the elastic resolution of the MARI spectrometer ($\Delta_\textrm{FWHM}=0.35$~meV for $E_\textrm{i} = 12.8$~meV), the observed excitations appear to be gapless at least within the instrumental resolution. We note that an additional peak develops at $|Q| = 1$~\AA$^{-1}$ ~below $T_\textrm{N}$, corresponding to the (1 1 1) the magnetic Bragg peak. This magnetic Bragg peak is consistent with the previous and our magnetic structure refinements using the propagation vector $\bm{k}=0$~\cite{Kawasaki}.

In order to calculate the spin wave dispersion, we construct the minimal spin Hamiltonian as follows,
\begin{equation} %Magnetic Hamiltonian
	\centering
	\mathcal{H} = J_1 \sum_{i} \hat{S}_i \cdot \hat{S}_{i+1} +J_2 \sum_{<i,j>} \hat{S}_{i} \cdot \hat{S}_{j} +J_3 \sum_{<i,j>} \hat{S}_{i} \cdot \hat{S}_{j+1},
	\label{Equation2}
\end{equation}
where $J_1$ is the intrachain interaction, $J_2$ and $J_3$ are the interchain couplings, $i$ and $j$ are the Mn site index on each chain and $\hat{S}$ are spin operators [see Fig.~\ref{Figure1}(b)]. Based on the spin Hamiltonian, we calculate the spin-spin correlation function, the neutron scattering cross-section, and the spin wave dispersions using linear spin-wave theory with the SpinW package~\cite{Toth}. The experimental resolution is included in the calculation. As a starting point, we simulate the spin wave dispersions based on the exchange interactions from DFT calculations~\cite{Chakraborty}.

\begin{figure}[h]
	\centering
	\includegraphics[width=1\linewidth, trim={0cm 90 0 70},clip]{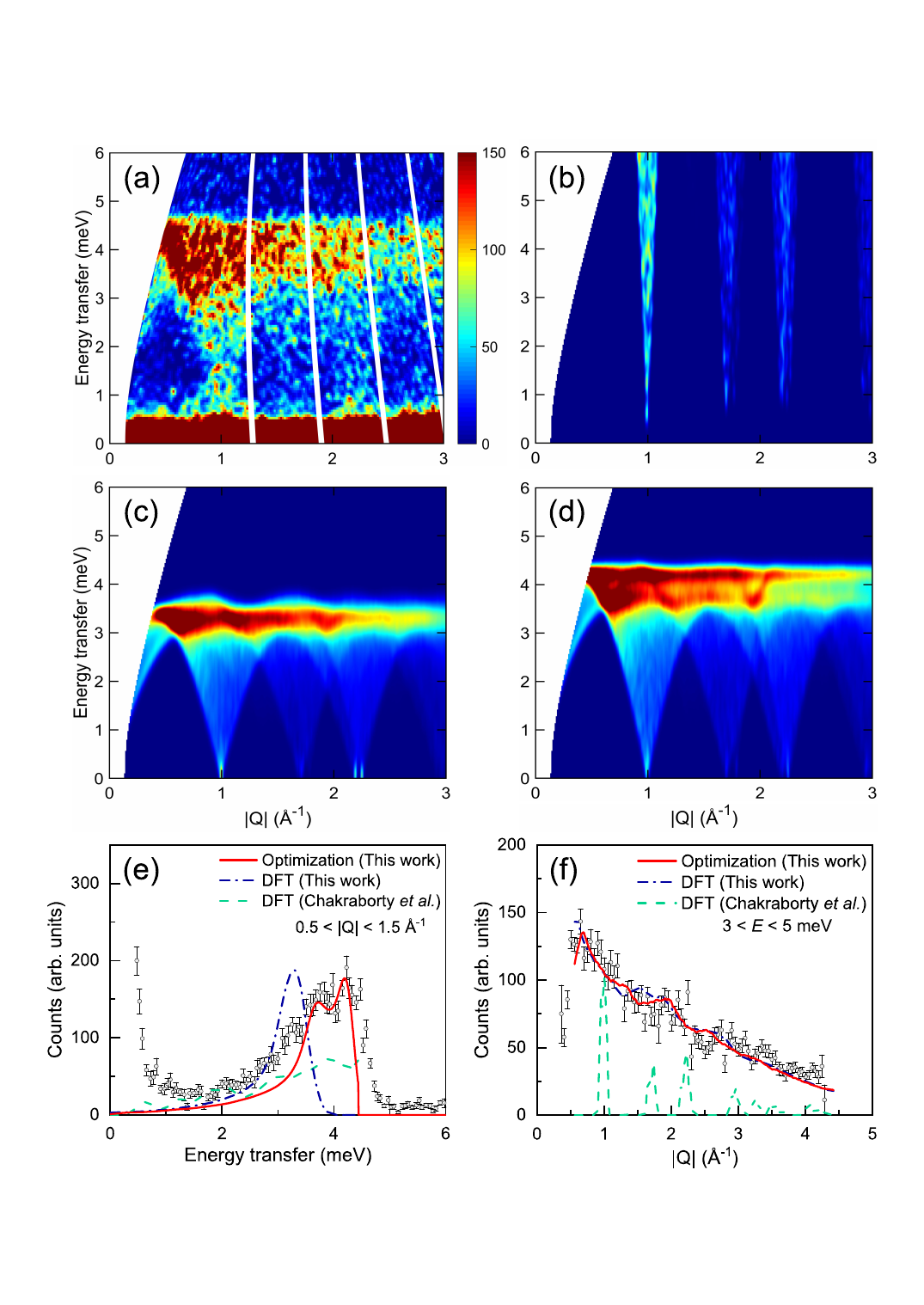}
	\caption{\textbf{Comparison with powder-averaged linear spin-wave theory calculations for different paramter sets.} (a) Magnetic excitations measured at $T = 5$ K. The calculated spin-wave excitations with (b) the previous DFT results in Ref.~\cite{Chakraborty} ($J_1 = 2.46$~meV, $J_2 = 1.76$~meV, and $J_3 = 0.34$ meV), (c) our DFT results ($J_1 = 0.716$~meV, $J_2 = 0.114$~meV, and $J_3 = -0.033$ meV), and (d) the optimized DFT values ($J_1 = 0.95$~meV, $J_2 = 0.15$~meV, and $J_3 = -0.01$~meV). One-dimensional cuts of the excitation spectrum at 5~K (e) integrated between $0.5 < |Q| < 1.5$~\AA$^{-1}$ and (f) between $3 < E < 5$~meV. The data are compared with three different calculation results.}
	\label{Figure7}
\end{figure}

As shown in Fig.~\ref{Figure7}(b), the calculations with the previous DFT exchange parameters in Ref.~\cite{Chakraborty} is incompatible with the measured excitation spectrum at $T = 5$~K, in which the simulated energy scale of the spin wave is five times higher than the data. Thus, the low-lying excitations are only observed within the energy range of the MARI spectrometer with $E_\textrm{i}= 12.8$~meV. This indicates that the given exchange parameter set is inappropriate to describe the observed magnetic excitations. In order to get insight into the energy scale of the exchange interactions, we evaluate the exchange coupling constants using a conventional DFT method. In this calculation, we first assume a ferromagnetic structure and compute the total energy. Next, based on the ferromagnetic state, we find five different magnetic sublattices by flipping each Mn spin. Then, we calculate the total energy for each magnetic state. By this means, we obtain the best results with the magnetic structure determined by our neutron diffraction measurements, in which the exchange coupling constants are evaluated to be $J_1 = 0.716$~meV, $J_2 = 0.114$~meV, and $J_3 = -0.033$ meV with the ordered moment of 2.91~$\mu_\textrm{B}$. 

The simulated spin-wave dispersions with our DFT results are broadly consistent with the measured excitation spectrum [see Fig.~\ref{Figure7}(c)]. However, a closer look reveals that the calculated spin waves are located at a slightly lower energy ($\sim 3.3$~meV). Furthermore, the flat spin-wave band is much narrower than the data, and the bumps appear on top of that. The differences between the data and calculations are clearly seen in Figs.~\ref{Figure7}(e) and \ref{Figure7}(f). The constant momentum cut for $0.5 < |Q| < 1.5$~\AA$^{-1}$ is inconsistent with the calculation based on the previous and our DFT results, as signified by different energy scales of the broad maximum. In addition, the constant energy cut for $3 < E < 5$~meV also manifests the deviation of the simulation from the experimental data. 

To reproduce the observed excitation with all main features, we optimize the exchange parameters obtained from the DFT calculations to $J_1 = 0.95$~meV, $J_2 = 0.15$~meV, and $J_3 = -0.01$~meV. With the optimized coupling constants, the magnetic excitations are well captured by the linear spin-wave theory, as shown in Fig.~\ref{Figure7}(d). Although the bandwidth of the calculated spin waves at around 4~meV is slightly narrower than the data [see Fig.~\ref{Figure7}(b)], the calculation with the optimized values reflects the main features of the excitation spectrum (Figs.~\ref{Figure7}(e) and \ref{Figure7}(f)). It should be noted that our calculation result exhibits no evidence for gapped magnetic excitations since we did not include in our Hamiltonian any terms accounting for anisotropy, whether Ising or easy-plane, originating from crystal fields effects or Dzyaloshinsky-Moriya interaction. However, considering the instrumental resolution ($\sim 0.35$~meV), we can not exclude the possibility of gapped low-lying excitations.

\begin{figure*}
	\centering
	\includegraphics[width=0.9\linewidth]{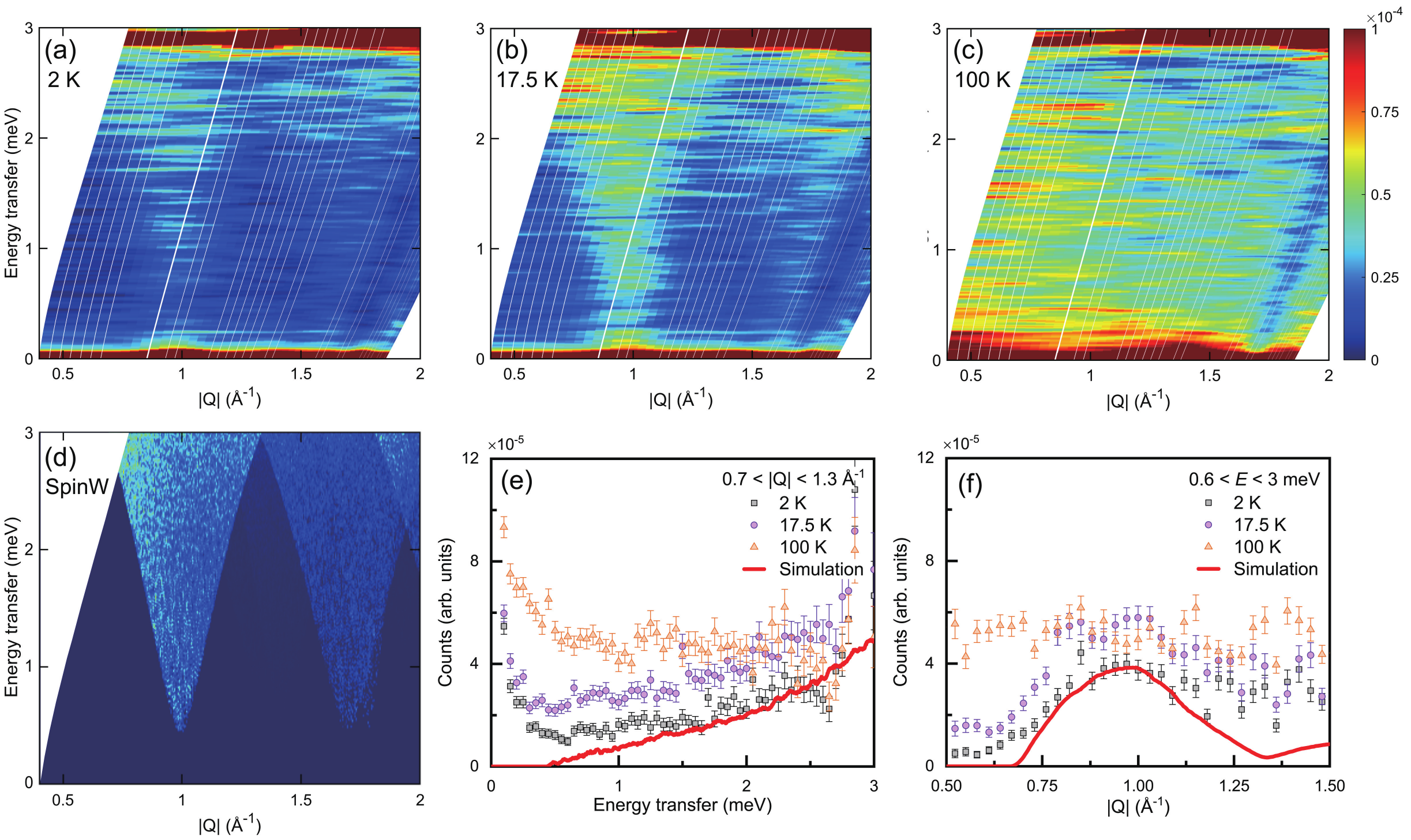}
	\caption{\textbf{Low-$E$ inelastic neutron scattering data and comparison with powder-averaged linear spin-wave theory calculations.} False color map of inelastic neutron scattering spectra measured at (a) 2~K, (b) 17.5~K, and (c) 100~K. (d) Powder-averaged calculated spin waves for the optimized exchange values with the single-ion anisotropy along the $a$-axis. One dimensional cuts of the magnetic excitations at various temperatures integrated for (e) $0.7 < |Q| < 1.3$~\AA$^{-1}$ and (f) $0.6 < E < 3$~meV.}
	\label{Figure8}	
\end{figure*}

To unveil the nature of low-lying magnetic excitations, we further performed INS experiments at the IRIS spectrometer. We summarize the low-$E$ INS results of Ca$_3$ZnMnO$_6$ at various temperatures in Fig.~\ref{Figure8}. At the lowest temperature $T = 2$~K, the spin gap is clearly observed with the gap size of about 0.5~meV [Fig.~\ref{Figure8}(a)]. The detected spin gap is comparable with the zero-field splitting observed in the high-frequency ESR measurements ($\sim0.69$~meV)~\cite{Ruan}. With increasing temperature, the spin gap gradually closes and completely disappears above $T_\textrm{N}$ in conjunction with the shift down of the spectral weight to the elastic scattering [Figs.~\ref{Figure8}(b) and \ref{Figure8}(c)]. 

The observed spin gap is well reproduced by introducing in the Hamiltonian a term corresponding to the single-ion anisotropy with the value of $-1.5$~$\mu$eV along the \textit{a}-axis [Fig.~\ref{Figure8}(d)]. As shown in Fig.~\ref{Figure8}(e), the constant momentum cut for $0.7 < |Q| < 1.3$~\AA$^{-1}$ shows a systematic decrease of the spectral weight near the elastic line. In addition, the constant energy cut for $0.6 < E < 3$~meV exhibits the predominant scattering intensity near 1~\AA$^{-1}$ below $T_\textrm{N}$ [Fig.~\ref{Figure8}(f)]. As illustrated by the red solid curve in Figs.~\ref{Figure8}(e) and ~\ref{Figure8}(f), our calculations using SpinW well reproduce the main features of the observed magnetic excitations at low $E$, including the low-$E$ spin-wave excitations and the opening of the spin gap $\Delta \sim 0.5$~meV. 

It should be noted that, for 3d transition-metal ions, magnetic anisotropy is generally negligible due to a relatively small spin-orbit coupling in an ideal octahedral crystal field~\cite{Abragam}. However, local symmetry lowering or local distortion of oxygen octahedra can induce single-ion anisotropy. In Ca$_3$ZnMnO$_6$, the small deviations from the ideal MnO$_6$ octahedra lift the degeneracy of the $t_{2g}$ orbital states as described in the next section. The crystal field effects due to the small distortions, in combination with the spin-orbit coupling, could give rise to the small magnetic anisotropy and consequently to the experimentally observed gap in the magnetic excitations.

\section{\label{sec4}Embedded dynamical mean field calculations (eDMFT)}

In materials where Mn$^{4+}$ ions occupy octahedral sites and exhibit a substantial charge gap, the fluctuating magnetic moment on Mn is typically temperature independent at $S=3/2$, originating from 
$t_{2g}$ electrons. As a result, the low-energy degrees of freedom are often well described by the Heisenberg Hamiltonian under a superexchange mechanism. Although this general picture mostly applies to
Ca$_3$ZnMnO$_6$, we show that, despite the predicted charge gap of about $2.1\,$eV, the superexchange is not the conventional type governed by virtual hopping of Mn–$t_{2g}$ electrons through oxygen. Instead, our theoretical calculations indicate that Mn–$e_g$ electrons are partially occupied—even though they do not contribute to the fluctuating local magnetic moment. These $e_g$ electrons are in a covalent bond with oxygen $p$ states via virtual hoppings, and also couple strongly to the $t_{2g}$ spins through Hund’s coupling. This scenario is reminiscent of the double-exchange mechanism found in colossal resistive manganites, yet here it occurs in a large-gap insulator. We further show that low-energy charge excitations also involve Zn–$d$
and Zn–$p$ states, and a substantial contribution from Ca–$p$ states. The Zn states mediate exchange along the chains, while the Ca–$p$ states govern exchange between the chains. Consequently, Ca$–p$
excitations and their coupling to the oxygens in Mn octahedra impart three-dimensional character to the spin exchange.

In this paper, calculations are performed using a state-of-the-art approach based on density functional theory combined with embedded dynamical mean-field theory (DFT+eDMFT) for correlated materials, as implemented in the eDMFT code~\cite{eDMFT1,eDMFT2,eDMFT3,eDMFT4}. Recent studies have demonstrated that DFT+eDMFT is predictive of both the electronic and structural properties of correlated materials at finite temperatures~\cite{Pred1,Pred2,Pred3,Pred4,Pred5,Pred6,Pred7}. This predictive capability has been showcased in other manganites, such as LaMnO$_3$ and BiMnO$_3$ (Mn$^{3+}$ ions)~\cite{Pred7}, Mn$_2$Mo$_3$O$_8$ (Mn$^{2+}$ ions)~\cite{Pred3,Pred5}, and PrBaMn$_2$O$_6$ (Mn$^{3.5+}$ ions)~\cite{Pred6,DPMn}, as well as in various other systems~\cite{Ni112,LK-99}.

A unique aspect of DFT+eDMFT is its use of quasi-atomic orbitals---which are more localized than Wannier orbitals---to describe the interacting part of the Hamiltonian, while the single-particle part is captured in the complete LAPW basis~\cite{eDMFT2,eDMFT3}. In addition, each correlated ion is described in a local coordinate system suitable for its specific geometry. For an ion in an octahedral environment, this local system is centered on the Mn ion, with the three axes directed toward the surrounding oxygen ions. In this coordinate system, the five $3d$ orbitals split into $t_{2g}$ and $e_g$ states. Because the octahedra in this material are slightly distorted---leading to angles that deviate from $90^\circ$---the $t_{2g}$ manifold further splits into a singlet $a_1$ and a doublet $e(1)$. The $e_g$ states are denoted by $e(2)$~\cite{Pred5}, as illustrated in the insets of Figs.~\ref{Figure_2T}(a)--(c).

\begin{figure*}
	\centering
	\includegraphics[width=0.9\linewidth]{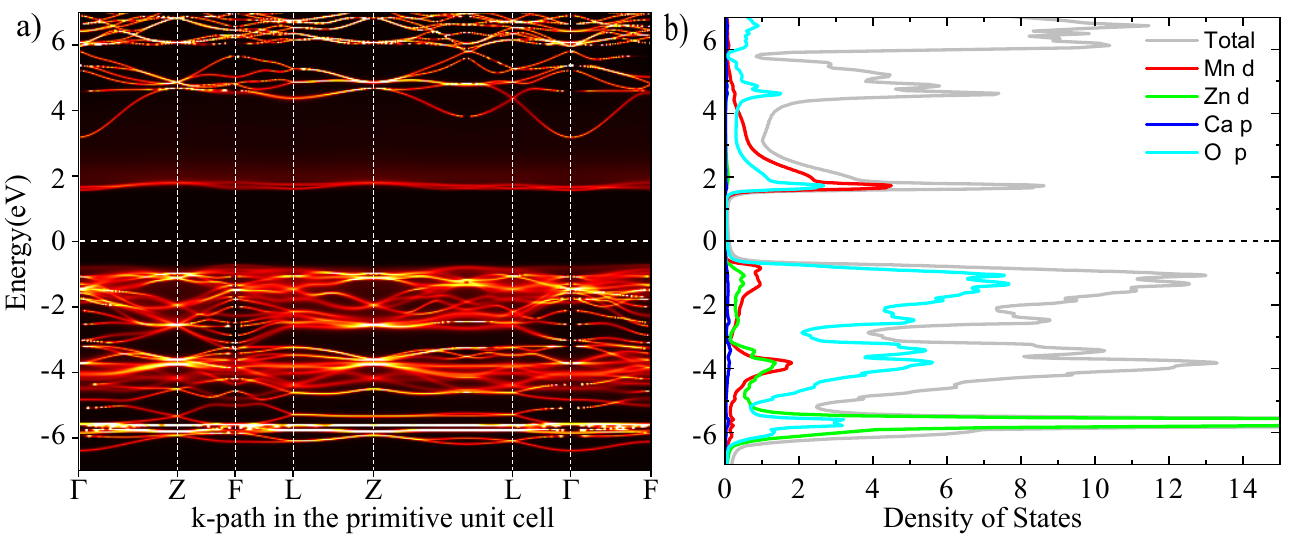}
	\caption{
    \textbf{Electronic properties of Ca$_3$ZnMnO$_6$ computed via the DFT+eDMFT method at 300\,K}. 
(a) Spectral function along the high symmetry path of the primitive cell;
(b) Total and orbital-projected density of states for the principal orbitals.   
}
\label{Figure_1T}
\end{figure*}
We start by presenting the Mn orbital occupancy, as calculated by real-space integration of the charge within the Mn muffin-tin sphere (which touches the oxygen muffin-tin sphere in the octahedra). While the expected occupancy is $n_d = 3$ electrons (based on the $\mathrm{Mn}^{4+}$ formal valence), our calculations reveal approximately $4.5$ electrons on the Mn ions, implying a valence of $\mathrm{Mn}^{2.5+}$. However, as we show below, only the $t_{2g}$ electrons are in a high-spin state and contributing to spin susceptibility, whereas the remaining $1.5$ electrons reside in $e_g$ states that are covalently coupled to the oxygen $p$ states.

To understand the charge excitations of Ca$_3$ZnMnO$_6$, we computed the single-particle spectral function along with the total and orbital-projected density of states (DOS), as shown in Fig.~\ref{Figure_1T}. The spectral function [Fig.~\ref{Figure_1T}(a)] exhibits both sharp bands, indicative of Bloch-like states, and incoherent spectral weight, suggesting the presence of local moments. The gap size is on the order of $\sim2.1$~eV, and the first particle and hole excitations originate primarily from hybridized Mn-$3d$ and O-$2p$ orbitals. Nevertheless, there is a substantial contribution from Zn-$d$ states, mainly in the occupied region of the spectrum, and a smaller (yet still significant) contribution from Ca-$p$ states that extend up to the valence-band edge, as also shown in Fig.~\ref{Figure_4T}(c). The first conduction band at $2\,$eV is flat in momentum and sharp in energy, corresponding to the antibonding branch of the covalent $e_g$--O-$p$ bond; the bonding branch is located at $-4\,$eV.

To understand the charge gap's complexity, we examine the orbital-projected DOS for the Mn-$3d$ orbitals and their corresponding self-energies (see Fig.~\ref{Figure_2T}). The behavior of the self-energy within the gap for a given orbital reveals the gap's origin, indicating whether it is driven by electronic correlations or covalent effects~\cite{Pred5,Pred7,LK-99}. For instance, if a given orbital's self-energy has a pole within the energy range of the gap, then the gap originates from electronic correlations (commonly termed a Mott gap). Conversely, if no pole is present and the imaginary part of the self-energy is zero within the gap, the gap arises from covalent effects or complete band filling, referred to as a band gap (when due to covalent effects, the orbital is partially occupied).
Our calculations show that the $t_{2g}$ orbitals (both $a$ and $e(1)$) have gaps opened by correlations (both being approximately half-occupied), while the $e(2)$ orbital has a gap opened by covalent effects and is partially occupied with considerable occupancy of $n_{e(2)}\approx 0.66$ (see Fig.~\ref{Figure_2T}). Electronic states exhibiting a mixture of Mott and band gaps are labeled in the literature as "orbital selective" states~\cite{Pred5}. However, the term "orbital selectivity" is more often used when at least one orbital remain metallic or semi-metallic~\cite{Pred7,OS1,OS2,OS3}.
   
\begin{figure*}
	\centering	\includegraphics[width=0.9\linewidth]{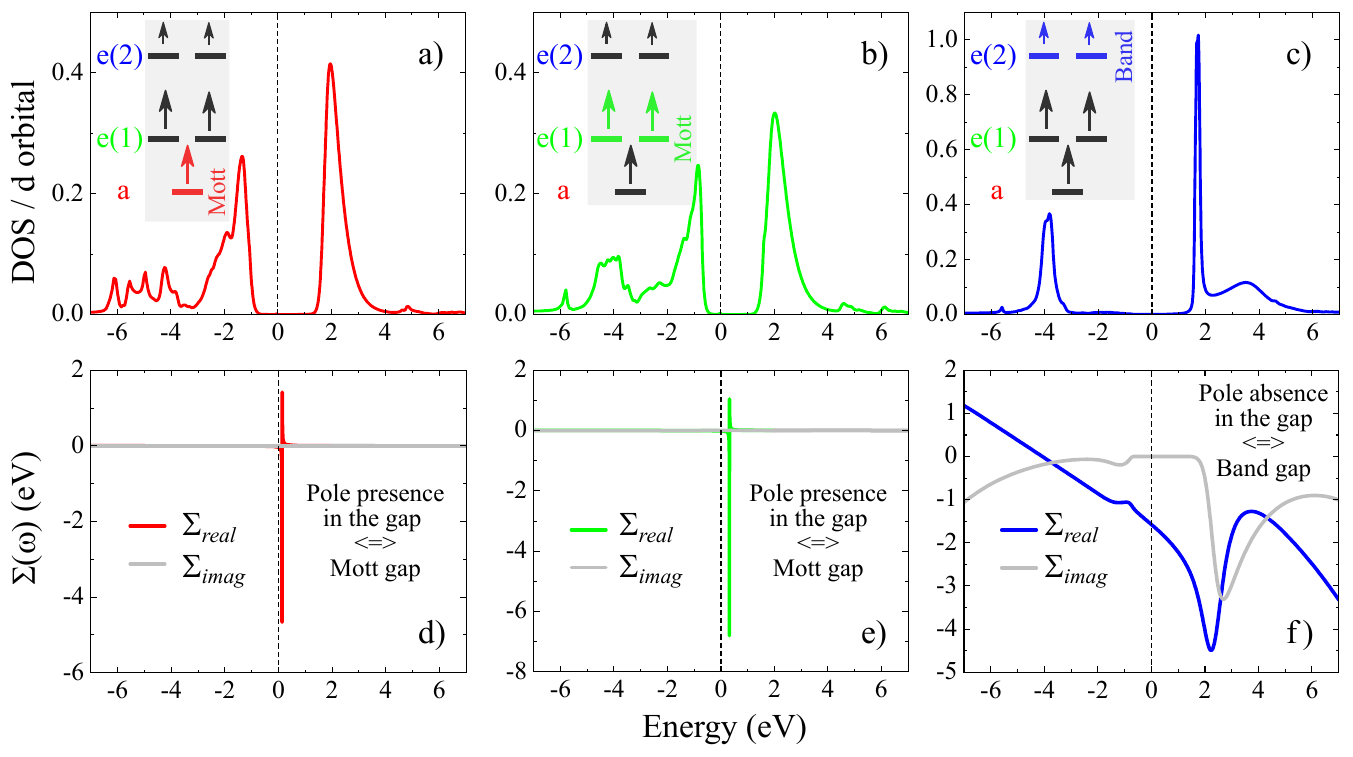}
	\caption{\textbf{Electronic properties of the Mn $3d$ orbitals in Ca$_3$ZnMnO$_6$ computed via the DFT+eDMFT method at 300\,K}. 
Panels~(a), (b), and (c) show the orbital-projected density of states (DOS), 
while panels~(d), (e), and (f) present the real and imaginary parts of the self-energy for the corresponding orbitals, respectively. 
The insets in (a)--(c) schematically depict the occupation of the three sets of nonequivalent orbitals: 
$a$ (a singlet state) and $e(1)$, $e(2)$ (doublet states). 
The arrow sizes represent the orbital occupation: 
the largest arrows in the $a$ and $e(1)$ orbitals indicate integer occupancy (one electron per orbital), 
whereas the smaller arrows in $e(2)$ reflect occupancy of less than one. 
Colors link each orbital to its DOS and self-energy: 
$a$~(red), $e(1)$~(green), and $e(2)$~(blue). 
The text insets in (d)--(f) describes how the shape of the self-energy identifies the gap type (Mott or band).}
\label{Figure_2T}
\end{figure*}
\begin{figure*}
	\centering
    \includegraphics[width=0.9\linewidth]{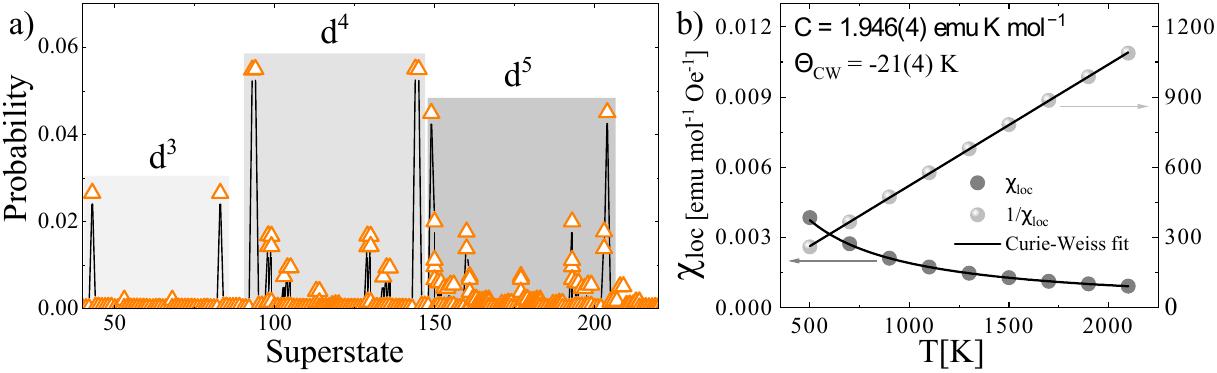}
	\caption{\textbf{Local properties of the Mn $d$ orbitals in Ca$_3$ZnMnO$_6$ computed via the DFT+eDMFT method at 300\,K}.
(a) Probability of the atomic superstates (orange triangles) for the Mn $3d$ orbitals, illustrating a mixture of $d^3$, $d^4$, and $d^5$ configurations. The solid line is a guide to the eye. 
Each superstate with finite probability corresponds to a particular orbital occupation and spin orientation (up or down). 
For example, the two most probable $d^3$ superstates are 
[$a$]$^\uparrow$[$e(1)$]$^{\uparrow\uparrow}$[$e(2)$]$^{00}$ and 
[$a$]$^\downarrow$[$e(1)$]$^{\downarrow\downarrow}$[$e(2)$]$^{00}$, 
whereas the four most probable $d^4$ superstates are 
[$a$]$^\uparrow$[$e(1)$]$^{\uparrow\uparrow}$[$e(2)$]$^{\uparrow 0}$, 
[$a$]$^\uparrow$[$e(1)$]$^{\uparrow\uparrow}$[$e(2)$]$^{0 \uparrow}$ (spin up), and 
[$a$]$^\downarrow$[$e(1)$]$^{\downarrow\downarrow}$[$e(2)$]$^{\downarrow 0}$, 
[$a$]$^\downarrow$[$e(1)$]$^{\downarrow\downarrow}$[$e(2)$]$^{0 \downarrow}$ (spin down), 
where $0$ denotes an empty orbital.
The $x$ axis enumerates all superstates:
number~1 corresponds to $d^0$,
numbers~2--11 to $d^1$,
12--42 to $d^2$,
43--92 to $d^3$,
and so on up to 352 for $d^{10}$. 
(b) Temperature dependence of the local magnetic susceptibility $\chi_\mathrm{loc}$ (dark circles) and its inverse $1/\chi_\mathrm{loc}$ (gray spheres). 
The solid lines are Curie--Weiss fits to the theoretical data, and the arrows indicate the respective $y$ axes for $\chi_\mathrm{loc}$ and $1/\chi_\mathrm{loc}$. 
The inset shows the Curie--Weiss constant $C$ and temperature $\Theta_\mathrm{CW}$ extracted from the fits.}
	\label{Figure_3T}
\end{figure*}
\begin{figure*}
	\centering
	\includegraphics[width=0.9\linewidth,trim={0cm 00 0 00},clip]{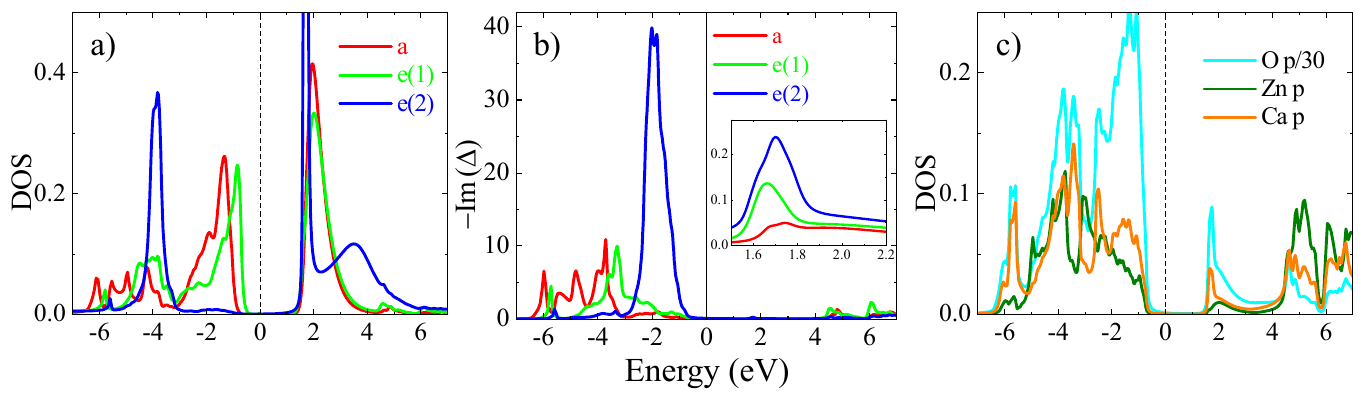}
	\caption{\textbf{Hybridization of the correlated orbitals with other states}. 
(a) Orbital-projected density of states (DOS) for the correlated orbitals, 
(b) imaginary part of the corresponding hybridization $\Delta(\omega)$, 
and (c) orbital-projected DOS for the noncorrelated ions. 
The inset in (b) magnifies the region in the conduction band where the $e(2)$ orbital exhibits a very narrow peak, arising from the covalent bond. 
}
	\label{Figure_4T}
\end{figure*}

To further investigate the origins of the orbital-selective state and the $d$-orbital occupation, we plot the probability of the atomic microstates for the Mn ion [see Fig.~\ref{Figure_3T}(a)]. The results indicate that the ion occupies a superposition of $d^3$, $d^4$, and $d^5$ configurations, with the highest probability for the $d^4$ configuration (corresponding to $\mathrm{Mn}^{3+}$), rather than the expected $d^3$ configuration (corresponding to $\mathrm{Mn}^{4+}$) based on charge neutrality.

To clarify this apparent discrepancy, we also computed the theoretical local magnetic susceptibility, shown in Fig.~\ref{Figure_3T}(b). The local magnetic susceptibility follows Curie--Weiss behavior, with an effective local moment of $3.91\,\mu_\mathrm{B}$, in good agreement with the experimentally obtained value of $3.96(5)\,\mu_\mathrm{B}$. The extracted theoretical Curie--Weiss temperature is somewhat smaller [$21(4)\,$K vs $41(5)\,$K] but remains consistent with the predominantly antiferromagnetic nature of the exchange interaction.

From the local magnetic susceptibility, we conclude that only the electrons occupying the Mott-insulating $t_{2g}$ orbitals contribute to the fluctuating or ordered magnetic moment. In contrast, the remaining $e_g$ electrons appear magnetically inert; however, they are not mere spectators. First, note that $t_{2g}$ and $e_g$ electrons are coupled by a strong Hund's coupling of about $1\,\mathrm{eV}$. Moreover, the $e_g$ electrons also couple strongly to oxygen $p$ electrons, as evidenced by the $6\,\mathrm{eV}$ bonding--antibonding splitting of the $e(2)$ orbital (bonding at $-4\,\mathrm{eV}$ and antibonding at $2\,\mathrm{eV}$, see Figs. Fig.~\ref{Figure_2T}(c) and Fig.~\ref{Figure_4T}(a)). Another way to understand how the different Mn orbitals interact with their surroundings is via the hybridization function shown in Fig.~\ref{Figure_4T}(b), which describes the strength of virtual hopping from an orbital to nearby states (here dominated by oxygen $p$ orbitals). We find that the $e(2)$ orbital exhibits a very large hybridization, with a peak at approximately $-2\,\mathrm{eV}$ reaching $40\,\mathrm{eV}$. By comparison, the $t_{2g}$ orbitals exhibit peaks below $-3$ and $-4\,\mathrm{eV}$ with values smaller than $10\,\mathrm{eV}$.
This result confirms the differing origins of the Mott gap and the band gap in these two orbital manifolds, while underscoring the strong covalent bonding between $e(2)$ and oxygen orbitals, which maintains large hybridization even at the valence-band edge. Consequently, the coupling of Mn orbitals to the surroundings is dominated by the $e(2)$--oxygen bond, so that the conventional superexchange mechanism (requiring sizable $t_{2g}$--oxygen hybridization near the valence-band edge) is weak. In contrast, a double-exchange mechanism---relying on Hund's coupling between $t_{2g}$ and $e_g$ orbitals, as well as strong $e_g$--oxygen coupling---clearly dominates in this material.

Because these exchange interactions are unconventional, our theoretical findings suggest that the distinct spin-exchange pattern observed in Ca$_3$ZnMnO$_6$ (as compared with other isostructural compounds) might stem from this double-exchange mechanism that necessitates orbital selectivity. Intriguing magnetic properties have been reported in other compounds that exhibit similar orbital selectivity~\cite{Pred3,Pred5,Pred6,Mag1,Mag2}, prompting us to speculate that systems with orbital-selective electronic states may display unexpected patterns of exchange interactions. Conversely, observing complex exchange patterns in certain materials could indicate the presence of novel, orbitally selective states.

For our theoretical calculations, we employed the experimental crystal structures listed in Table~\ref{Table1}. The Kohn--Sham potential was obtained via the generalized gradient approximation using the Perdew--Burke--Ernzerhof (GGA-PBE) functional~\cite{PBE}. The LAPW basis was defined with a plane-wave cutoff of $\texttt{RKmax} = 7.0$, and we used $1024$ \textbf{k}~points in the irreducible part of the Brillouin zone.
We used the hybridization-expansion quantum impurity solver (CT-HYB)~\cite{ctqmc} with a fully rotationally invariant Coulomb interaction. The eDMFT functional was evaluated using the exact double-counting scheme~\cite{exactd}, assuming dielectric screening of the Coulomb repulsion in real space. To construct the DMFT projector, we employed quasi-atomic orbitals (solutions of the radial Schrödinger equation linearized at the Fermi level), choosing the local coordinate axes according to the local geometry. These orbitals were embedded in a wide hybridization window from $-10$\,eV to $+10$\,eV relative to the Fermi level. The partially screened Coulomb interactions were set at $U = 10$\,eV and $J_\mathrm{H} = 1$\,eV, values well established in the literature for insulating transition-metal compounds~\cite{Pred3,Pred5,Pred7,SM1,SM2,SM3}. The spin-orbit coupling was neglected in this calculation.
The self-energy on the real axis was obtained via the maximum entropy analytical continuation of the local cumulant, as detailed in Ref.~\cite{eDMFT1}.
The local magnetic susceptibility is computed by the quantum impurity solver sampling $\chi_S(i\omega_n)=\int_0^\beta \exp(i\omega_n\tau)\langle{S_z(\tau)S_z(0)}\rangle$, and converting it to the cgs unit system, emu$\cdot$mol$^{-1}\cdot$Oe$^{-1}$, using the formula $\chi_\textrm{loc}^\textrm{cgs}=129.26~10^{-6}~\chi_S(\omega=0)$.

The local effective moment can be computed using the following formula 
$\mu_\textrm{eff}^\textrm{loc}=2.827\sqrt[]{\chi_\textrm{loc}^\textrm{cgs}~\times~T[\textrm{K}]}$~\cite{CWchi}.

\section{\label{sec5}Discussion and Conclusion}

Our characterization results reveal that Ca$_3$ZnMnO$_6$ hosts 3D long-range magnetic order, which is contrary to the expected quasi-1D spin topology. As the intrachain coupling $J_1$ is much larger than the interchain couplings $J_2$ and $J_3$, it is naturally expected to observe the characteristics of quasi-1D magnetism. However, the absence of a broad maximum in $C_\textrm{m}$ at high $T$ and the observed magnetic excitations defy the 1D nature of a ground state. Such disagreement can be understood in terms of the energy hierarchy of exchange interactions, which could be a consequence of the unconventional double-exchange-like mechanism found by the DFT+eDMFT method. The exchange coupling constants of Ca$_3$ZnMnO$_6$ are determined to be $J_1 = 0.95$~meV $J_2 = 0.15$~meV, and $J_3 = -0.01$~meV. Compared with the isostructural compound Ca$_3$Co$_2$O$_6$ ($J_1\sim0.5$~meV, $J_2\sim0$~meV, and $J_3\sim0.05$~meV)~\cite{Paddison2014}, Ca$_3$ZnMnO$_6$ has non-negligible value of $J_2$, possibly playing a significant role in the 3D long-range ordering. 

We turn to the spin dimensionality of Ca$_3$ZnMnO$_6$. As evidenced by our NPD results and the previous high-frequency ESR results~\cite{Ruan}, the Ising nature of the 3d moments observed in Ca$_3$Co$_2$O$_6$ is not preserved for Ca$_3$ZnMnO$_6$. A recent inelastic X-ray scattering study on Ca$_3$Co$_2$O$_6$ has proposed that the origin of the Ising-like anisotropy is a consequence of the high-spin Co$^{3+}$ ($d^6$) ions situated at the trigonal prism site (B site)~\cite{Leedahl2019}. Since in Ca$_3$ZnMnO$_6$, the B sites are occupied by nonmagnetic Zn ions, prismatic trigonal crystal field can not give rise to any anisotropy, thus we are left with the easy-plane anisotropy at the octahedral site due to the crystal field and spin-obit coupling effects.

\begin{figure}[h]
	\centering
	\includegraphics[width=1\linewidth, trim={0cm 270 0 270},clip]{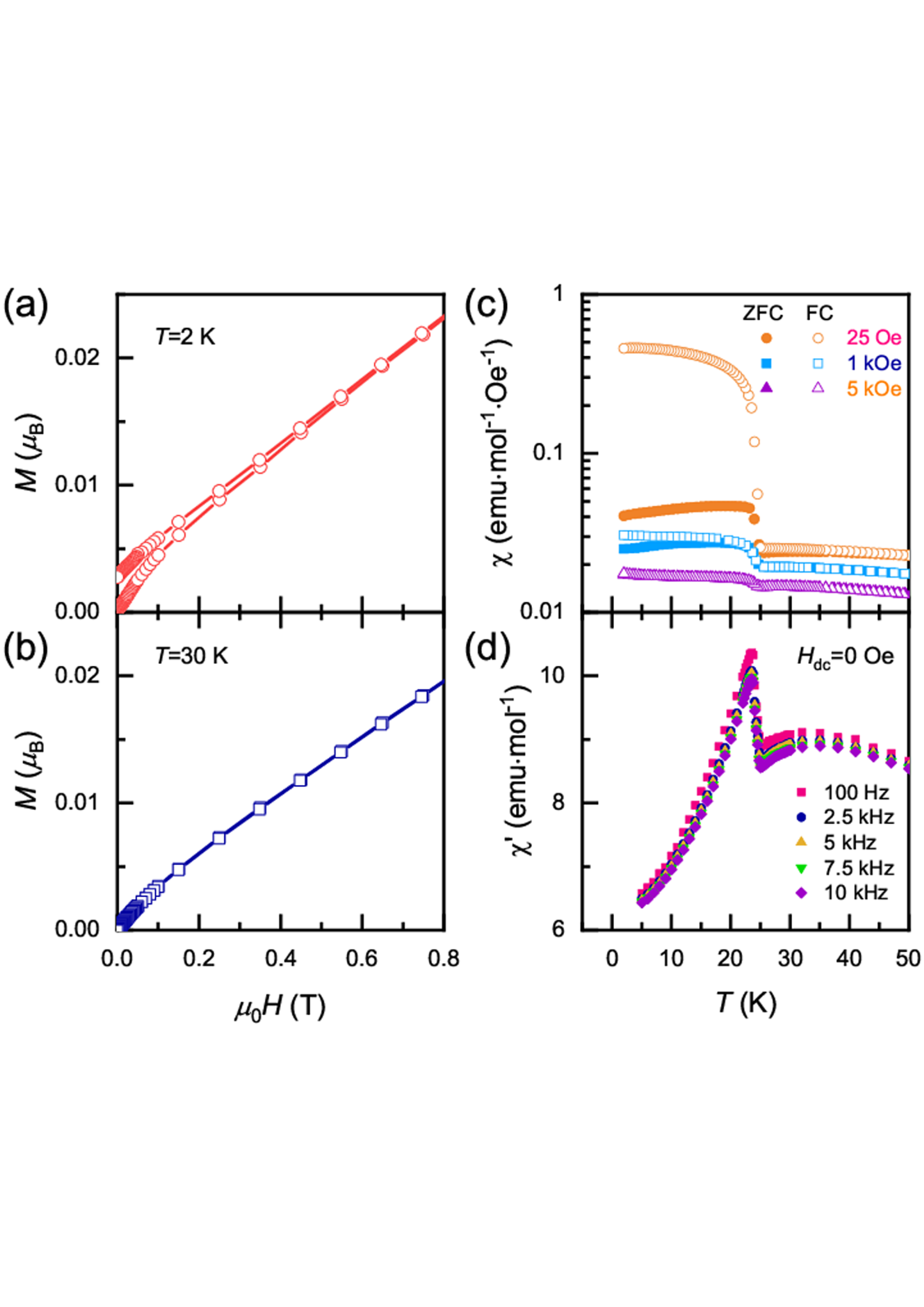}
	\caption{\textbf{Magnetic characterization results of the polycrystalline Ca$_3$ZnMnO$_6$} (a),(b) Isothermal magnetization curves at 2 and 30~K. (c) Field dependence of the dc magnetic susceptibility in a semilog scale. (d) Frequency dependence of the ac magnetic susceptibility in the frequency range of $0.1-10$~kHz.}
	\label{Figure13}
\end{figure}

Lastly, we discuss the possibility of altermangeitsm in Ca$_3$ZnMnO$_6$. Figures~\ref{Figure13}(a) and \ref{Figure13}(b) present the isothermal magnetization curves at 2 and 30~K. At 2~K, $M(H)$ exhibits field hysteresis up to 0.8~T, while $M(H)$ at 30~K shows a convex behavior at low fields and linear increment at higher fields. The observed hysteresis behavior below $T_\textrm{N}$ indicates the presence of ferromagnetic contributions, further corroborated by the suppression of the ZFC-FC divergence in $\chi(T)$ with increasing field [Fig.~\ref{Figure13}(c)]. It is well established that altermagnets with collinear antiferromagnetic arrangements preserve a symmetry that facilitates ferromagnetic behaviors, giving rise to non-zero net magnetic moments~\cite{Cheong2024-1,Cheong2024-2}. One might expect the presence of partial spin-glass behavior in the magnetically ordered state. However, this possibility can be ruled out because of the absence of frequency dependence in the ac magnetic susceptibility $\chi'(T)$ [Fig.~\ref{Figure13}(d)].

\begin{figure}[h]
	\centering
	\includegraphics[width=0.7\linewidth]{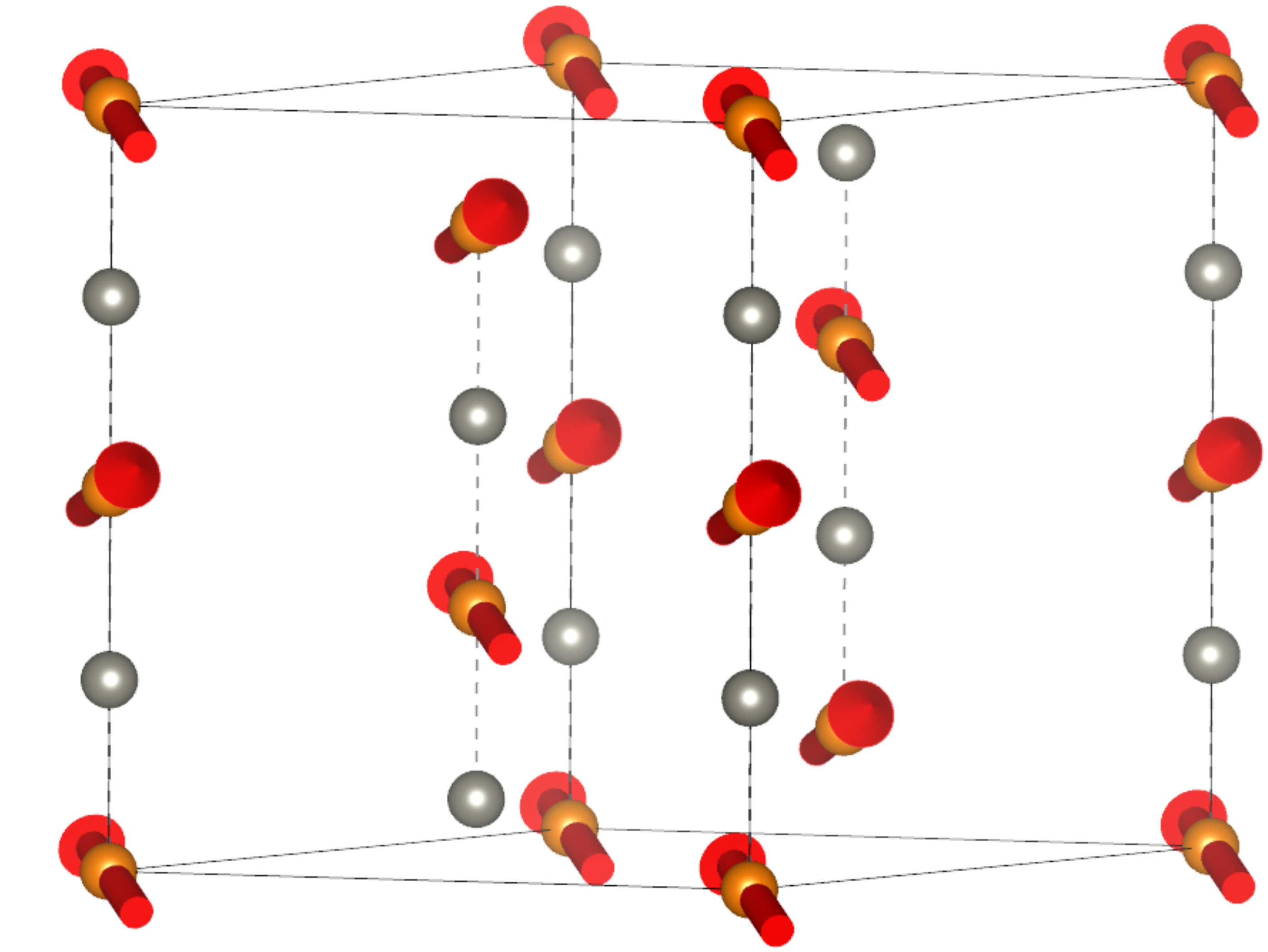}
	\caption{\textbf{Magnetic model, consistent with M-type altermagnetism}, inferred from neutron scattering, magnetization, and high-field ESR; the red arrows represent the magnetic moments, while the orange and gray spheres represent the Mn and Zn ions, respectively.}
	\label{FigureT5}
\end{figure}

To further strengthen the possibility of altermagnetism, we have performed symmetry analyses of the magnetic structure,  taking into account the paramagnetic symmetry, the easy-plane anisotropy demonstrated by high-field ESR~\cite{Ruan}, as well as neutron diffraction and magnetization measurements (the latter indicate a small ferromagnetic component, see Fig. \ref{Figure13}). Although high-field ESR together with powder neutron diffraction measurements reveal that the main components of the magnetic moments are close to or lie in the ab-plane of the paramagnetic crystal structure, powder neutron diffraction is not sensitive to the small ferromagnetic component found in the magnetization measurements. Combining these findings together, we propose a slightly different model for the magnetic structure, shown in Fig.~\ref{FigureT5}. This magnetic model can be described either by the P$\bar{1}$ or C2$’$/c$’$ magnetic space groups with $\bar{1}$ and 2$’$/m$’$ magnetic point groups, respectively. In the C2$’$/c$’$ moments are aligned along the two-fold axis of the monoclinic space group. Both magnetic point groups are consistent with the possibility of altermagnetism~\cite{Cheong2024-1,Cheong2024-2,Paolo}.We hypothesize that the tiny ferromagnetic component, arising from the crystal field combined with spin-orbit coupling~\cite{Ruan}, is aligned along the c-axis. According to the classification of altermagnets for non-collinear spins, our system satisfies the requirements for an M-type altermagnet~\cite{Cheong2024-1,Cheong2024-2,Paolo}. Single-crystal neutron diffraction studies are necessary to prove the model shown in Fig.~\ref{FigureT5} and distinguish between the two magnetic space groups, P$\bar{1}$ and C2$’$/c$’$.

To conclude, we have explored the spin dynamics and magnetic excitations of the quasi-1D spin chain Ca$_3$ZnMnO$_6$ using basic characterization, $\mu$SR, NPD, and INS. Ca$_3$ZnMnO$_6$ exhibits a commensurate antiferromagnetic order at $T_\textrm{N}=25$~K, in which the Mn spins lie in the $ab$-plane. In contrast to the expected properties due to the low dimensionality of the magnetic ions, quasi-1D spin chain, we find signatures of 3D magnetic order, evinced by typical 3D gapped dispersive spin wave excitations below $T_\textrm{N}$. 
Nonetheless, modeling the data using the linear spin-wave theory gives rise to the energy hierarchy corresponding to a quasi-1D spin system. 
Such discrepancy is attributed to the enhanced interchain interaction $J_2$ by Zn substitution into the trigonal prism site, which leads to unconventional superexchanges with orbital selectivity. In addition, we speculate that materials with orbital-selectivity may show unexpected exchange patterns or vice versa complex exchange patterns could indicate the presence of novel states with orbital selectivity.  

\section*{Acknowledgments}
S.L. acknowledges support from the Institute for Basic Science (IBS-R014-Y2). D. T. A. thank Engineering and Physical Sciences Research Council, UK for funding (Grant No. EP/W00562X/1) and the CAS for PIFI Fellowship. D. T. A. would like to thank the Royal Society of London for International Exchange funding between the United Kingdom and Japan and Newton Advanced Fellowship funding between the United Kingdom and China. K.H. work was supported by NSF DMR-2233892.
G.L.P.’s work was supported by a grant of the Romanian
Ministry of Education and Research, CNCS
- UEFISCDI, Project No. PN-III-P1-1.1-TE-2019-1767,
within PNCDI III. We thank ISIS Facility for beam time on GEM [https://doi.org/10.5286/ISIS.E.24003413], on MARI [https://doi.org/10.5286/ISIS.E.5512195, https://doi.org/10.5286/ISIS.E.47622999] on IRIS [https://doi.org/10.5286/ISIS.E.24072759] and MuSR [RB1010488]. Computing resources for the theoretical calculations were provided by STFC Scientific Computing Department’s SCARF cluster. Preparation of the input files and data processing for the theoretical calculations were performed on the local cluster at Stefan cel Mare University of Suceava (USV), obtained through a grant of the Romanian Ministry of Education and Research, CNCS—UEFISCDI, project number PN-III-P1-1.1-TE-2019-1767, within PNCDI III. We thank John Taylor for help on MARI experiment and Stefano Agrestini, Geetha Balakrishnan and Vivek Kumar Anand for help on the sample preparation. S.W.C. was supported by the W. M. Keck foundation grant to the Keck Center for Quantum Magnetism at Rutgers University. The work at SKKU was supported by the National Research Foundation (NRF) of Korea (Grants No. RS-2023-00209121 and No.
2020R1A5A1016518).

\end{document}